\documentclass[prd,aps,preprintnumbers,showpacs,amsfonts,amsmath,floatfix,nofootinbib]{revtex4}
\usepackage{latexsym,graphicx,epsfig,psfrag,here,bm,natbib}
\newcommand{\spur}[1]{\!\not\! #1}
\newcommand{\bra}[1]{ \left |  #1 \right >}
\renewcommand{\bra}[1]{ \left |  #1 \right >}
\newcommand{\qm}{|\vec q|}
\newcommand{\qv}{{\vec q}\,}

\newcommand{\Mvariable}{}
\newcommand{\CapitalIota}{I}
\newcommand{\CapitalEpsilon}{E}
\newcommand{\Mfunction}{}
\newcommand{\dd}{\displaystyle}

\begin{document}

\preprint{DSF-2006-21 (Napoli)}

\title{Semileptonic B decays into even parity charmed mesons}

\author{Manuela De Vito} \author{Pietro Santorelli}
\affiliation{Dipartimento di Scienze Fisiche, Universit{\`a} di
Napoli ``Federico II", Italy,\\
Istituto Nazionale di Fisica Nucleare, Sezione di Napoli, Italy}

\begin{abstract}
\noindent
By using a constituent quark model we compute the form factors relevant to
semileptonic transitions of B mesons into low-lying $p$-wave charmed mesons.
We evaluate the $q^2$ dependence of these form factors and compare them
with other model calculations. The Isgur-Wise functions $\tau_{1/2}$  and
$\tau_{3/2}$ are also obtained in the heavy quark limit of our results.
\end{abstract}

\maketitle

\section{Introduction}
\label{s:intro}

Recently, BaBar Collaboration has discovered a narrow state with
$J^P=0^+$ with a mass of 2317 MeV, $D_{s0}^\ast(2317)$
\cite{BabarDs}. The existence of a second narrow resonance, $D_{sJ}(2460)$ with $J^P=1^+$,
was confirmed by CLEO \cite{CleoDs}. Both states have been confirmed by BELLE
\cite{BelleDs}. Soon after the discovery, another set of charmed
mesons, $D_0^{\ast 0}(2308)$ and $D_1^{\prime 0}(2427)$ which have
the same quantum numbers $J^P=(0^+,1^+)$ as $D_{sJ}$ has been
discovered by BELLE \cite{BelleD}. Before their discovery, quark
model and lattice calculations predicted that the masses of these
states, in particular $D_{s0}^\ast(2317)$ and $D_{s1}^\prime(2460)$,
would be significantly higher than observed \cite{Isgur},\cite{lattice}.
Moreover, these states were predicted to be broad due to the fact that they
can decay into $D\, K$ and $D^\ast \ K$, respectively. Experimentally, the masses of
$D_{s0}^\ast(2317)$ and $D_{s1}^\prime(2460)$ are below the $D\, K$ and $D^\ast \ K$
thresholds and hence they are very narrow.  These facts inspired a lot of theorists
to explain the puzzle \cite{teorici}.

In this paper we will focus our attention on the weak semileptonic
transitions of $B$ mesons into lower lying $p$-wave charmed mesons ($D^{\ast\ast}$).
These transitions were studied, within a quark model approach, for the first time in
\cite{ISGW} and, more recently, in \cite{QuarkModels} where the authors
take into account the symmetries of QCD for heavy quarks \cite{HQET},
already used in \cite{IsgurWiseEP}. The light-front covariant model \cite{Jaus}
was adopted to study the same subject in \cite{ChengEP}. The
relevant form factors were also evaluated, in the framework of QCD
Sum Rules \cite{QCDSR}, in \cite{QCDSR_EP}.

Here we employ a simple constituent quark model \cite{ioetalbari,noi}
to evaluate semileptonic form factors of B mesons into $p$-wave
charmed mesons. The plan of the paper is the following. In the next section we
describe our quark model; the third section is devoted to introduce and evaluate
the $s$-wave to $p$-wave form factors. Our way to fix the free parameters of
the model and the resulting form factors are discussed in section four, while in
section five the heavy quark limit of the form factors are computed and
compared with Heavy Quark Effective Theory predictions; the $\tau_{1/2}$, and $\tau_{3/2}$
are also evaluated. In the last section we show and discuss our numerical results.

\section{A Constituent Quark Model}
\label{s:model}

\begin{figure}[t]
\begin{center}
\epsfig{file=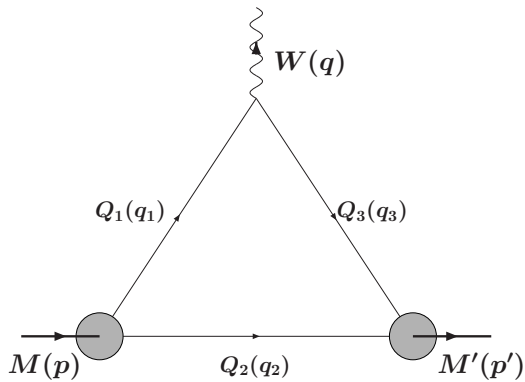,height=5cm}
\end{center}
\vspace{-0.4truecm}
\caption{\label{f:trBD}
Quark model diagram for the hadronic
amplitude relevant to semileptonic $M\to M^\prime$ decay
involving $Q_1\to Q_3$ transitions. The thin lines represent
quarks, the thick ones mesons. The gray disks represent the
quark-quark-meson vertexes.}
\end{figure}

\noindent
In our model \cite{ioetalbari,noi} any heavy meson
$H(Q \overline{q})$, with $Q\in\{b,c\}$ and $q\in\{u,d,s\}$, is described by the matrix
\begin{equation}
H=\frac{1}{\sqrt 3}\psi_H (k)\;\; \frac{\spur{q_1}+m_{Q}}{2
m_{Q}}\, \Gamma\, \frac{-\!\!\!\spur{q_2}+m_{q}}{2 m_{q}}\,,
\label{e:Hi}
\end{equation}
where $m_{Q}$ ($m_q$) stands for the heavy (light) quark mass;
$q^\mu_1,\ q^\mu_2$ are their $4-$momenta (cfr Fig.~\ref{f:trBD}).
$\psi_H(k)$ indicates the meson's wave function which is fixed by
using a phenomenological approach. The meson constituent quarks
vertexes, $\Gamma$ in Eq. (\ref{e:Hi}), are fixed by using the
correct transformation properties under C, P and to enforce the relation
\begin{eqnarray}
<H|H>& \equiv & {\rm Tr}\{(-\gamma_0 H^\dagger\gamma_0)\ H\}\ =
\int \frac{d^3 k}{(2\pi)^3} |\psi_H(k)|^2 =  2\ M_H\, .
\label{e:funnorm}
\end{eqnarray}

\noindent For the odd parity, $s$-wave heavy mesons $J^P=(0^-,1^-)$,
$\Gamma$ is given by
\begin{eqnarray}
\Gamma_{0^-} & = &  -\imath \gamma_5 {\sqrt{ \frac{2\, m_{Q}
m_{q}}{m_H^2 - (m_{Q}- m_{q})^2} }}
\;,  \\
\Gamma_{1^-}(\varepsilon,q_1,q_2) & = &
\varepsilon_\mu\left[\gamma^\mu-\frac{q_1^\mu-q_2^\mu}{m_H+m_{Q}+m_{q}}\right]
{\sqrt{ \frac{2\, m_{Q} m_{q}}{m_H^2 - (m_{Q}- m_{q})^2} }} \;,
\end{eqnarray}
where $\varepsilon$ is the polarization $4-$vector of the (vector)
meson $H$.
The vertexes of the lower lying even parity heavy mesons, instead, are given by the matrices
\footnote{We don't consider in this paper the tensor mesons.}
\begin{eqnarray}
\Gamma_{^3\!P_0} = \Gamma_{0^+} & = &  -i  \sqrt{\frac{2\, m_Q m_q}{m_H^2- (m_Q + m_q)^2}}\;,  \\
\Gamma_{^3\!P_1}(\varepsilon,q_1,q_2) & = &
\varepsilon_\mu\left[\gamma^\mu-\frac{(m_Q-m_q)(q_1^\mu-q_2^\mu)}{-m_H^2+(m_Q-m_q)^2}\right]
\gamma_5 \sqrt{\frac{3\, m_Q m_q}{m_H^2- (m_Q + m_q)^2}}\;,  \\
\Gamma_{^1\!P_1}(\varepsilon,q_1,q_2) & = &
\varepsilon_\mu\left[\frac{(q_1^\mu-q_2^\mu)\;
m_H}{m_H^2-(m_{Q}-m_{q})^2}\right] \gamma_5 \sqrt{\frac{6\; m_Q
m_q}{m_H^2-(m_Q+m_q)^2}}\;.
\end{eqnarray}
As already discussed in \cite{ioetalbari,noi}, the $4-$momentum
conservation in the meson-constituent quarks vertexes
can be obtained defining a heavy running quark mass.
For details we address the reader to the references \cite{ioetalbari,noi}.
Here, for the sake of utility, we recall all the remaining rules of our
model for the evaluation of the hadronic matrix elements of weak
currents:
\begin{itemize}
\item[a)] for each quark loop with $4$-momentum $k$ we have
\begin{equation}
\int\frac{d^3k}{(2\pi)^3}\;\; , \label{e:loop}
\end{equation}
a colour factor of 3 and a trace over Dirac matrices;
\item[b)] for the weak hadronic current, $\overline{q}_2\
\Gamma^\mu\ q_1$, one puts the factor
\begin{equation}
\sqrt{\frac{m_{1}}{E_{1}}}\ \sqrt{\frac{m_{2}}{E_{2}}}\
\Gamma^\mu\;, \label{e:J}
\end{equation}
where with $\Gamma^\mu$ we indicate a combination of Dirac
matrices.
\end{itemize}

\section{Form Factors}
\label{s:DCFDF}

In this section we evaluate the form factors parameterizing the
$0^-\to (0^-,1^-)$ and $0^-\to (0^+,{^3\!P_1},{^1\!P_1})$
weak transitions. The decomposition of these matrix elements
of weak currents in terms of form factors are the following (see also \cite{ISGW})
\begin{eqnarray}
<H^\prime_{0^-}(p^\prime)|V_{\mu} |H_{0^-}(p)> & = &
f_{+}(q^2)(p_\mu+p_\mu^\prime) + f_{-}(q^2)(p_\mu-p_\mu^\prime)\;,
\label{e:F1F0}\\
<H^\prime_{1^-}(p^\prime,\varepsilon)|V_{\mu}-A_{\mu} |H_{0^-}(p)>
& = &  2\ g(q^2)\ \epsilon^{\mu\nu\alpha\beta}\
\varepsilon^\ast_\nu\, p_\alpha\, p^\prime_\beta \nonumber \\
& & -\imath \left\{f(q^2)\, \varepsilon^\ast_\mu\ +
(\varepsilon^\ast \cdot p)\, \left
[\frac{}{}a_+(q^2)\,(p_\mu+p^\prime_\mu)+a_-(q^2)\,(p_\mu-p^\prime_\mu)\right]
\right\}\;,
\label{e:gfapam}\\
<H_{0^+}(p^\prime)|A_{\mu} |H_{0^-}(p)> & = &
F_{+}(q^2)(p_\mu+p_\mu^\prime) + F_{-}(q^2)(p_\mu-p_\mu^\prime)\;,
\label{e:F1F00p}\\
<H_{^3\!P_1}(p^\prime,\varepsilon)|V_{\mu}-A_{\mu} |H_{0^-}(p)> &
= & -\imath \left\{F^\prime(q^2)\, \varepsilon^\ast_\mu\ +
(\varepsilon^\ast \cdot p)\, \left
[\frac{}{}A^\prime_+(q^2)\,(p_\mu+p^\prime_\mu)+A^\prime_-(q^2)\,(p_\mu-p^\prime_\mu)\right]
\right\}\nonumber\\
&&+2\ G^\prime(q^2)\ \epsilon^{\mu\nu\alpha\beta}\
\varepsilon^\ast_\nu\, p_\alpha\, p^\prime_\beta \;,
\label{e:gfapam3P1}\\
<H_{^1\!P_1}(p^\prime,\varepsilon)|V_{\mu}-A_{\mu} |H_{0^-}(p)> &
= & -\imath \left\{F(q^2)\, \varepsilon^\ast_\mu\ +
(\varepsilon^\ast \cdot p)\, \left
[\frac{}{}A_+(q^2)\,(p_\mu+p^\prime_\mu)+A_-(q^2)\,(p_\mu-p^\prime_\mu)\right]
\right\}\nonumber\\
&&+2\ G(q^2)\ \epsilon^{\mu\nu\alpha\beta}\ \varepsilon^\ast_\nu\,
p_\alpha\, p^\prime_\beta \;. \label{e:gfapam1P1}
\end{eqnarray}
The calculation of the form factors in Eqs. (\ref{e:F1F0}) and
(\ref{e:gfapam}) for the case of $B\to D(D^\ast)$ transitions has
been done in Ref. \cite{noi}. However, for the sake of utility,
the analytical expressions are reported in appendix \ref{a:0m0m1m}.\\
One of the main results of this paper is the calculation of the
form factors appearing in Eqs. (\ref{e:F1F00p}),
(\ref{e:gfapam3P1}) and (\ref{e:gfapam1P1}). By way of an example,
in the following we describe the calculation of the matrix element
$<H_{0^+}(p^\prime)|A_{\mu} |H_{0^-}(p)>$ and give the expressions
of the form factors $F_\pm$. In appendix \ref{a:0m1p} we collect the
expressions for $G^{(\prime)}$, $F^{(\prime)}$ and
$A^{(\prime)}_\pm$. Note that all the calculations are done in the frame where
$q^\mu_3 = (E_3,\vec{k}-\vec{q})$,
$q^\mu_1 = (E_1,\vec{k})$, $q^\mu_2 = (E_2,-\vec{k})$.

Let us start considering the $0^- \to 0^+$
transition,
\begin{eqnarray}
<H_{0^+}(p^{\prime})|\bar Q_3 \gamma_{\mu}\gamma_5 Q_1|H_{0^-}(p)>
& = & - \int_{\cal
D}\frac{d^3k}{(2\pi)^3}\psi^\ast_{0^+}(k)\psi_{0^-}(k)
\sqrt{\frac{m_1 m_3}{E_1 E_3}}\nonumber\\
&& {\rm Tr}\left[ \frac{-\spur{q_2}+m_2}{2
m_2}(\gamma_0\Gamma_{0^+}(q_3,q_2)^\dagger\gamma_0) \frac{\spur{q_3}+ m_3}{2 m_3}\
\gamma_\mu\gamma_5 \frac{\spur{q_1}+m_1}{2 m_1}(\Gamma_{0^-})
\frac{-\spur{q_2}+ m_2}{2 m_2} \right ],
\label{e:0m0p}
\end{eqnarray}
where ${\cal D}$ is the integration domain (see Refs.
\cite{ioetalbari,noi}) defined by
\begin{eqnarray}
{\rm Max}(0,k_{-})\leq & k & \leq {\rm Min}(K_M,k_{+})\nonumber\\
{\rm Max}(-1,f(k,|\qv|))\leq & cos(\theta) & \leq +1 \\
0\leq & \phi & \leq 2 \pi \nonumber \label{e:domain}
\end{eqnarray}
with
\begin{eqnarray}
k_{\pm} & = & \frac{ \qm \, (m_{F}^2+m_2^2)\pm
(m_{F}^2-m_2^2) \sqrt{m_{F}^2+\qv^2}}{2 m_{F}^2}\, ,\\
f(k,\qm) & = & \frac{2 \sqrt{m_{F}^2+\qv^2}
\sqrt{k^2+m_2^2}-(m_{F}^2+m_2^2)}{2k\,\qm}\, .
\label{e:f}
\end{eqnarray}
$\phi$  and $\theta$ are the azimuthal and the polar angles respectively
for the tri-momentum $\vec k$. $K_M=(m_I^2-m_2^2)/(2m_I)$ and $m_I$ ($m_F$)
is the mass of the initial (final) meson: in Eq. (\ref{e:0m0p}) $m_F=m_{0^+}$.
We choose the $z-$axis along the direction of $\qv$, the (tri-)momentum of the W
boson (cfr Fig. \ref{f:trBD}).

The analytical expressions for the form factors can be obtained by comparing Eq. (\ref{e:0m0p})
with Eq. (\ref{e:F1F00p})
\begin{eqnarray}
\label{e:F1F00pE-a}
F_+(q^2)  & = &  \int  \frac{k^2 dk d
cos\theta \psi_I(k)\psi_F^\ast(k)}{8\,{\pi }^2\,{m_{\CapitalIota }}\,
  {\sqrt{\left( -{{d_{12}}}^2 + {{m_{\CapitalIota }}}^2 \right) \,
      \left( {{m_F}}^2 - {{s_{23}}}^2 \right) \,{{\CapitalEpsilon }_1}\,
      {{\CapitalEpsilon }_3}}}}\times \nonumber\\
&&\left[ {d_{12}}{s_{23}}{{\CapitalEpsilon }_2} -
  {m_{\CapitalIota }}\left( \left( {d_{13}} - 2{m_2} \right) {m_2} +
     {m_{\CapitalIota }}{{\CapitalEpsilon }_2} \right)+
     \frac{k}{\qm}\cos (\theta )\left( - {{{\Mfunction{m}}_F}}^2{m_{\CapitalIota }}
         + {{\Mfunction{d}}_{12}}{s_{23}}
     \left( {m_{\CapitalIota }} - {{\CapitalEpsilon }_F} \right)  +
    {{{\Mfunction{m}}_{\CapitalIota }}}^2{{\CapitalEpsilon }_F} \right)
     \right]\,,\\
\label{e:F1F00pE-b}
F_-(q^2)  & = &  \int  \frac{k^2 dk d
cos\theta \psi_I(k)\psi_F^\ast(k)}{8\,{\pi }^2\,{m_{\CapitalIota }}\,
  {\sqrt{\left( -{{d_{12}}}^2 + {{m_{\CapitalIota }}}^2 \right) \,
      \left( {{m_F}}^2 - {{s_{23}}}^2 \right) \,{{\CapitalEpsilon }_1}\,
      {{\CapitalEpsilon }_3}}}}\times \nonumber\\
&& \left[ {m_2}{m_{\CapitalIota }}{s_{13}} + \left(
{{m_{\CapitalIota }}}^2 + {d_{12}}{s_{23}} \right)
{{\CapitalEpsilon }_2} -
\frac{k}{\qm}  \cos (\theta )
\left( {{{\Mfunction{m}}_F}}^2{m_{\CapitalIota }} +
{{\Mfunction{d}}_{12}}{s_{23}}\left({m_{\CapitalIota }}+ {{\Mfunction{\CapitalEpsilon }}_F}\right) +
      {{{\Mfunction{m}}_{\CapitalIota }}}^2 {{\Mfunction{\CapitalEpsilon }}_F}
    \right)\right]\,,
\end{eqnarray}
where ($d_{ij} = m_i-m_j$ and $s_{ij} = m_i+m_j$).

The expressions for the remaining form factors in Eqs.
(\ref{e:F1F0})-(\ref{e:gfapam1P1}) are collected in appendix \ref{a:0m0m1m} and \ref{a:0m1p}.

\section{Fixing the free Parameters}
\label{s:freepar}

The numerical evaluation of the form factors
given in Section \ref{s:DCFDF} requires to specify the
expression for the vertex functions and the values of the free
parameters of the model. For the vertex functions we adopt two
possible forms, the gaussian-type, extensively used in literature
(see for example \cite{ioemisha})
\begin{equation}
\psi_H(k) = 4\,\sqrt{ M_H}\,
\sqrt{\frac{\sqrt{\pi^3}}{\omega_H^3}}\,
\exp\left\{-\frac{k^2}{2\, \omega_H^2}\right\}\, ,
\label{e:wfgauss}
\end{equation}
and the exponential one
\begin{equation}
\psi_H(k) = 4 \pi \sqrt{\frac{M_H}{\omega_H^3}}\,
\exp\left\{-\frac{k}{\omega_H}\right\}\, , \label{e:wfexp}
\end{equation}
which is able to fit the results of a relativistic quark model
regarding the shape of the meson wave-functions \cite{salpeternardulli}. In our
approach $\omega_H$ is a free parameter which should be fixed by
comparing a set of experimental data with the predictions of the
model. In this paper we choose to fix the free parameters by a fit
to the experimental data on the $Br(B\to D \ell \nu)$
\cite{PDG} and on the spectrum of $B\to D^\ast \ell \nu$ process \cite{BDstar}.

\begin{figure}[t]
\begin{center}
\epsfig{file=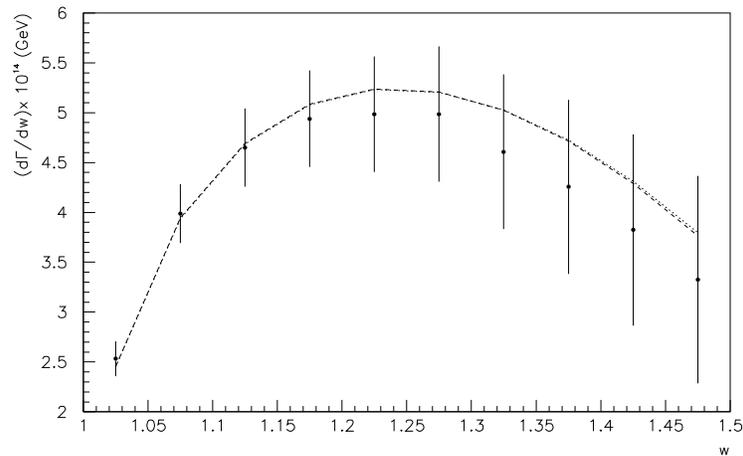,height=6cm}
\end{center}
\vspace{-0.5truecm}
\caption{\label{f:spettro} The $B\to D^\ast \ell \nu_\ell$ spectrum.
The dashed (dotted) line corresponds
to the exponential (gaussian) vertex function. The data are taken from Ref. \cite{BDstar} }
\end{figure}

The quality of the agreement between fitted spectrum and the corresponding
experimental data may be assessed by looking at the Figure \ref{f:spettro}.
It should be also observed the very small differences between the
$B\to D^\ast \ell \nu$ spectrum using the vertex functions in
Eqs. (\ref{e:wfgauss})-(\ref{e:wfexp}). Regarding the $B\to D \ell \nu$ branching ratio,
we obtain $2.00\, (2.01)\,  \% $
for the exponential (gaussian) vertex function to be compared to the experimental value:
$ Br(B\to D \ell \nu)\ =\  2.15\ \pm\ 0.22\ (2.12\ \pm\ 0.20) \% $ for the
charged (neutral) B meson.
At this stage the two different form of the vertex functions agree equally well with
the experimental data. However, differences emerge when single form factors are considered
(cfr for example Table \ref{t:fdfris1}).

\begin{center}
\begin{table}[t!!!]
\caption{The two best fit sets of values of
the free parameters are obtained for the exponential (exp.) and
gaussian (gauss.) vertex function. \label{t:datafit1}}
\vspace{-0.2truecm}
\begin{tabular}{|c|c|c|}
\hline
Parameter & fitted values (exp.) & fitted values (gauss.)    \\
\hline\hline
$m_q$     & ~23 MeV~ & ~34 MeV~ \\
\hline
$\omega_B$     & ~101 MeV~ & ~108 MeV~ \\
\hline
$\omega_D=\omega_{D^\ast}=\omega_{D^{\ast\ast}}$     & ~51 MeV~ & ~186 MeV~ \\
\hline
$|V_{cb}|$     & ~0.041~ & ~0.043~ \\
\hline
\end{tabular}
\end{table}
\end{center}

\section{Heavy Quark Limit}
\label{s:HQL}
In this section we perform the heavy quark limit for the
form factors obtained in the previous sections. Before to do this
we briefly remind the implications of the HQET on the heavy meson spectrum.
In the quark model, mesons are conventionally classified according to
the eigenvalues of the observables $J$, $L$ and $S$: any state is
labelled with the symbol $^{2S+1}L_J$. So, if we consider the lower lying
even parity mesons ($L=1$), the scalar and the tensor mesons correspond to
$^3\!P_0$ and $^3\!P_2$ states, respectively. Moreover, there are two
states with $J=1$: the $^1\!P_1$ and $^3\!P_1$, they can mix each other if the
constituent quark masses are different as in the case of charmed
mesons. For heavy mesons the decoupling of the spin of
the constituent heavy quark, $s_Q$, suggests to use a different
set of observables: the total angular momentum of the light
constituent, $j_q (=s_q+L)$, the orbital momentum of the light
degree of freedoms respect to the heavy quark, $L$, the total
angular momentum $J\, (=j_q+s_Q)$, any state is labelled with
$L^{j_q}_J$. In this representation the scalar  and the tensor
mesons are labelled with $P^{1/2}_0$ and $P^{3/2}_2$,
respectively. The axial mesons are classified as $P^{3\!/2}_1$ and
$P^{1\!/2}_1$ and are related to the ${^3\!P_1}$ and ${^1\!P_1}$
states by the \cite{IsgurWiseEP}
\begin{eqnarray}
\label{e:1p13p1HQa}
\bra{P^{3\!/2}_1} & = & + \sqrt{\frac{2}{3}} \left | {^1\!P_1}\right > + \sqrt{\frac{1}{3}} \bra{^3\!P_1}\;, \\
\label{e:1p13p1HQb}
\bra{P^{1\!/2}_1} & = & + \sqrt{\frac{1}{3}} \bra{^1\!P_1} -
\sqrt{\frac{2}{3}} \bra{^3\!P_1}\;.
\end{eqnarray}
The scaling laws of the HQET concern the $0^-\to (P^{1/2}_0\,,P^{1\!/2}_1\,,P^{3\!/2}_1\,)$ form factors,
they are defined combining the form factors in Eqs. (\ref{e:F1F00p}), (\ref{e:gfapam3P1}),
(\ref{e:gfapam1P1}) and the relations in Eqs. (\ref{e:1p13p1HQa}) and (\ref{e:1p13p1HQb}).
For example $G^{1\!/2}(q^2)\equiv \sqrt{1/3}\,G(q^2)-\sqrt{2/3}\,G^\prime(q^2)$.

To extract the heavy quark mass dependence from the expressions of the
form factors we follow the same approach used in our previous paper \cite{noi}.
We introduce the variable $x$, defined by  $x= (2 \alpha k)/m_F$, in such a
way, neglecting the light quark mass respect to the heavy ones, the integration
domain, near the zero-recoil point, simplify to
$0\leq x \leq \alpha$, $0\leq \theta\leq\pi$ and $0\leq \phi\leq 2 \pi$.
Therefore, if we look at the expressions of $F_\pm(q^2)$, Eqs. (\ref{e:F1F00pE-a})-(\ref{e:F1F00pE-b}),
near the zero recoil point (i. e. $q^2\simeq q^2_{max}$)
we have, neglecting terms of the order of $x^3$,
\begin{equation}
\left . F_\pm(q^2)\frac{}{}\right |_{q^2\simeq q^2_{max}} =
\int_0^{\alpha}\; dx\; \psi_I(k(x))\; \psi_F(k(x))\
\frac{ m_F^2\,\left( m_F \mp m_I \right) \,\left( 7 - 3\,w \right) \,x^2  }
{384\,m_I\,{\pi }^2\,{\alpha }^3}\, .
\label{e:fpfmHQL}
\end{equation}
For $\alpha\ll 1$ the integration can be easily done giving
\begin{equation}
F_\pm(q^2)|_{q^2 \simeq q^2_{max}} =
\frac{m_F \mp m_I}{\sqrt{m_F m_I}} \left\{
\begin{array}{lcl}
\dd \left [ 2\sqrt{2}
\left(\frac{\omega_I\, \omega_F}{\omega_I^2+\omega_F^2}\right)^{3/2} \right ]
\frac{1}{3}\left( 1-\frac{3}{4}(w-1) \right ) + o((w-1)^2)
&  & {\rm gaussian-type}\\ \\
\dd \left [ 8
\frac{\sqrt{\omega_I^3 \omega_F^3}}{(\omega_I+\omega_F)^3}\right ]
\frac{1}{3}\left( 1-\frac{3}{4}(w-1) \right )+ o((w-1)^2)
& & {\rm exponential-type}\,.
\end{array}\right.
\label{e:FpFmHQL}
\end{equation}
Thus, the $\tau_{1/2}$ Isgur-Wise function resulting from our model is given by
\begin{equation}
\tau_{1/2}(w) = \frac{1}{3}-\frac{1}{4}(w-1) + \frac{19}{96} (w-1)^2 + o((w-1)^3)\,,
\end{equation}
where we have also written the term $(w-1)^2$ which was neglected in
Eq. (\ref{e:FpFmHQL}).

A similar analysis can be performed on the heavy to heavy
$0^- \to P^{1\!/2}_1$ form factors.
Let us start considering the combination
\begin{equation}
\left\{
\begin{array}{c}
G^{1\!/2}(q^2)\\
F^{1\!/2}(q^2)\\
A^{1\!/2}_\pm(q^2)
\end{array}
\right\}
=
\frac{1}{\sqrt{3}}\left\{
\begin{array}{c}
G(q^2)\\
F(q^2)\\
A_\pm(q^2)
\end{array}
\right\}
-
\sqrt{\frac{2}{3}}\left\{
\begin{array}{c}
G^\prime(q^2)\\
F^\prime(q^2)\\
A^\prime_\pm(q^2)
\end{array}
\right\}
\end{equation}
which defines the $0^- \to P^{1\!/2}_1$ form factors. It is very
simple to obtain their scaling laws in the limit of heavy quark
masses. Following the above method we obtain
\begin{equation}
\left\{
\begin{array}{c}
G^{1\!/2}(q^2)\\
F^{1\!/2}(q^2)\\
A^{1\!/2}_\pm(q^2)
\end{array}
\right\}
=
N\dd\left\{
\begin{array}{l}
\sqrt{m_I m_F}\\
2\sqrt{m_I m_F}(w-1)\\
\dd\mp 1/\sqrt{m_I m_F}
\end{array}
\right\}
\frac{1}{3}\left( 1-\frac{3}{4}(w-1)
\right ) + o((w-1)^2)\,,
\end{equation}
where
\begin{equation}
N=
\left\{
\begin{array}{ll}
\dd  2\sqrt{2} \left(\frac{\omega_I\, \omega_F}{\omega_I^2+\omega_F^2}\right)^{3/2} & {\rm gaussian-type}\\
& \\
\dd  8 \frac{\sqrt{\omega_I^3 \omega_F^3}}{(\omega_I+\omega_F)^3} &  {\rm exponential-type} \;.
\end{array}
\right.
\end{equation}

Similarly, we can evaluate the $\tau_{3/2}$ Isgur-Wise function obtaining
\begin{equation}
\tau_{3/2}(w) = \frac{5}{6}-\frac{31}{24}(w-1) + \frac{93}{64} (w-1)^2  + o((w-1)^3)\;.
\end{equation}
A comparison between our results and some others
coming from quark models, QCD sum rules and Lattice calculations can be done
looking at the Table \ref{t:IW}. The values of the $\tau$ functions at zero
recoil point and their slopes are compatible. In particular, it
should be observed that our results for $\tau_{1/2}$ are practically the same obtained in the
Isgur Scora Grinstein Wise (ISGW) model \cite{ISGW} and QCD Sum Rules findings
\cite{DaiHuang,ColangeloDeFazioPaver}.%
\footnote{The values and the slopes
of $\tau$ functions are obtained fitting the numerical results obtained in
Morenas {\it et al.} (see Ref \cite{QuarkModels}) using the ISGW model.}
Regarding $\tau_{3/2}$, our result at zero recoil point is
slightly larger of the results coming from other models, while the slope
is comparable with others.

Relations between the slope of Isgur-Wise function and $\tau$
functions at zero recoil points were derived, in the form of sum rules, by Bjorken
\cite{Bjorken} and Uraltsev \cite{Uraltsev}
\begin{eqnarray}
\rho^2 -\frac{1}{4} & = & \sum_n |\tau^{(n)}_{1/2}(1)|^2 + 2\sum_n |\tau^{(n)}_{3/2}(1)|^2\, ,\\
\frac{1}{4} & = & \sum_n |\tau^{(n)}_{3/2}(1)|^2 - \sum_n |\tau^{(n)}_{1/2}(1)|^2\, ,
\end{eqnarray}
where $n$ stands for the radial excitations and
$\rho^2$ is the slope of the Isgur-Wise function  $\xi(w)$ which, in our model \cite{noi}, is
\begin{equation}
\xi(w) = 1-\frac{11}{12}(w-1) + \frac{77}{96}(w-1)^2 +
o((w-1)^3)\, .
\end{equation}
Our results for $n=0$ oversaturate both the sum rules. For the Bjorken sum rule this
is due to the small value we obtain for the derivative of the Isgur-Wise function
($\rho^2$) which is in any case compatible with the experimental value
$\rho^2 = 0.95\pm 0.09$ \cite{BDstar}. We plan to study this problem in a separate work.
However, a detailed discussion on these sum rules and the findings of quark models
can be found in \cite{Memorino}.

\begin{table}[t]
\caption{\small{\label{t:IW} The Isgur-Wise functions $\tau_{1/2}$ and $\tau_{3/2}$ at
zero recoil and their slope parameters.}}
\vspace{-0.3truecm}
\begin{center}
\begin{tabular}{||c|c|c|c|c||}
\hline\hline
$\tau_{1/2}(1)$ & $\rho_{1/2}^2$ & $\tau_{3/2}(1)$ & $\rho_{3/2}^2$ & Ref. \\
\hline\hline
 0.33 & 0.75 & 0.83 & 1.29 & This work \\
 0.34 & 0.76 & 0.59 & 1.09 & \cite{ISGW}\\
 0.22 & 0.83 & 0.54 & 1.5  & \cite{BTfrancesi}\\
 0.31 & 1.18 & 0.61 & 1.73 & \cite{Cheng} \\
 $0.13 \pm 0.04$ & $0.50\pm 0.05 $ & $0.43 \pm 0.09$ & $0.90\pm0.05$ & QCDSR \cite{DaiHuang}  \\
 $0.35\pm0.08$ & $2.5\pm1.0$ & -- & -- & QCDSR(NLO)\cite{ColangeloDeFazioPaver} \\
 $0.38 \pm 0.04$  &  & $0.53 \pm 0.08 $ &  &  Lattice \cite{tauLattice}\\
 \hline
\end{tabular}
\end{center}
\end{table}

\section{Numerical results and discussion}
\label{s:results}

\begin{table}
\caption{\small{
\label{t:fdfris1}
$B\to D^{\ast\ast}$ form factors evaluated at $q^2 =0$ and at $q^2_{max} = (m_B-m_{D^{\ast\ast}})^2$
by using the vertex function in Eq. (\ref{e:wfexp}). In parentheses
the values obtained using the gaussian vertex function (cfr. Eq. (\ref{e:wfgauss})).
}}
\vspace{-0.3truecm}
\begin{center}
\begin{tabular}{|c|c|c|c|}
   \hline
Form Factor & This work  & Ref. \cite{ISGW} &  Ref. \cite{ChengEP} \\
\hline
    & F(0) \hspace{3truecm} \hfill{F($q^2_{max}$)} & F(0)
\hspace{0.5truecm} \hfill{F($q^2_{max}$)} &
      F(0) \hspace{0.5truecm} \hfill{F($q^2_{max}$)} \\
\hline\hline
$F_1$ & -0.32\;(-0.30)  \hfill{-0.35\;(-0.33)} & -0.18 \hfill{-0.24} & -0.24 \hfill{-0.34} \\
\hline
$F_0$ & -0.32\;(-0.30) \hfill{-0.025\;(-0.036)} & -0.18 \hfill{0.008} & -0.24 \hfill{-0.20} \\
\hline
$A_{0}^{(1/2)}$ & -0.25\;(-0.23) \hfill{-0.31\;(-0.28)} & -0.18 \hfill {-0.39} & -0.075 \hfill{-0.083} \\
\hline
$A_{1}^{(1/2)}$ & 0.096\;(0.088) \hfill{-0.0018\;(-0.0029)} & 0.070 \hfill{-0.002} & 0.073 \hfill{0.071} \\
\hline
$A_{2}^{(1/2)}$ & 0.69\;(0.63) \hfill{0.87\;(0.79)} & 0.49 \hfill{0.91} & 0.32 \hfill{0.56} \\
\hline
$V_{}^{(1/2)}$ & 0.67\;(0.61) \hfill{0.84\;(0.76)} & 0.44 \hfill{0.81} & 0.31 \hfill{0.55} \\
\hline
$A_{0}^{(3/2)}$ & -0.61\;(-0.58) \hfill{-0.81\;(-0.77)} & -0.20 \hfill{-0.46} & -0.47 \hfill{-0.76} \\
\hline
$A_{1}^{(3/2)}$ & -0.13\;(-0.13) \hfill{-0.016\;(-0.023)} & -0.005 \hfill{-0.008} & -0.20 \hfill{-0.26} \\
\hline
$A_{2}^{(3/2)}$ & 0.70\;(0.65) \hfill{1.27\;(1.19)} & 0.33 \hfill{0.72} & 0.25 \hfill{0.47} \\
\hline
$V_{}^{(3/2)}$ & -0.81\;(-0.77) \hfill{-1.09\;(-1.03)} & -0.44 \hfill{-0.71} & -0.61 \hfill{-1.24} \\
\hline
\end{tabular}
\end{center}
\end{table}

All the results discussed in the previous section has been obtained without fixing the
free parameters of the model. In this one we use the fitted values of the free
parameters in Table \ref{t:datafit1} (cfr section \ref{s:freepar} for discussion) to obtain
the numerical results of the form factors. First of all, in Table \ref{t:fdfris1} are
collected the values of the form factors for the $B\to D^{**}$ transitions
evaluated at zero recoil point ($q^2_{max}=(m_B-m_{D^{\ast\ast}})^2$) and at $q^2=0$. We consider the
charmed final state with the following masses: $m(D_0^*) = 2.40\; GeV$,
$m(D_1^\prime) = 2.43\; GeV$, $m(D_1) = 2.42\; GeV$ \cite{PDG}.%
\footnote{$D_1^\prime$ and $D_1$ represent, respectively, the two different physical
axial-vector charmed meson states. The physical $D_1^\prime$ ( $D_1$) is
primarily  $P_1^{1/2}$ ($P_1^{3/2}$). They differ by a small amount from the mass
eigenstates in the heavy quark limit, for a discussion see \cite{ChengProduction}.
In this paper we neglect these differences. }
Note that we are considering, for a better comparison with other calculations, the elicity form
factors (cfr for definitions, for example, \cite{Lusignoli}). Looking at the Table \ref{t:fdfris1}, we can see
that the absolute values of our form factors (at $q^2=0$ ) are larger than
the ones in Ref \cite{ISGW,ChengEP}, this naturally implies larger branching
ratios in our model. In particular, our predictions on the branching ratios,
using the exponential (gaussian)
vertex function, are ($\tau_{B^0}= 1.536\, \times\, 10^{-12}$ s \cite{PDG})
\begin{eqnarray}
Br(\bar B^0\to D^{*+}_{0}     \ell^- \bar\nu_\ell) & = & 2.3 (2.1) \times 10^{-3} \left (|V_{cb}|/0.041\right)^2\,, \nonumber\\
Br(\bar B^0\to D^{\prime +}_1 \ell^- \bar\nu_\ell) & = & 2.0 (1.6) \times 10^{-3} (|V_{cb}|/0.041)^2\,, \\
Br(\bar B^0\to D^{+}_1        \ell^- \bar\nu_\ell) & = & 8.0 (7.3) \times 10^{-3} (|V_{cb}|/0.041)^2\,. \nonumber
\end{eqnarray}

\begin{table}
\caption{\small{
\label{t:fdfris2}
Parameters of the $B\to D^{\ast\ast}$ form factors. The functional $q^2$ dependence is either
polar : $F(q^2) = F(0)/(1-a\, q^2/m_B^2)$ or linear: $F(q^2) = F(0)(1+ b\, q^2)$. In parentheses
the values obtained using the gaussian vertex function (cfr. Eq. (\ref{e:wfgauss})). }}
\vspace{-0.3truecm}
\begin{center}
\begin{tabular}{|c|c|c|c|}
   \hline
Form Factor & F(0)  & $a$  & $b\ (GeV^{-2})$ \\
\hline\hline
$F_1$ & -0.32 (0.30) & 0.263 (0.233)  &     \\
\hline
$F_0$ & -0.32 (0.30) &   & -0.109 (-0.103)   \\
\hline
$A_{0}^{(1/2)}$ & -0.25 (-0.23) & 0.661 (0.626) &    \\
\hline
$A_{1}^{(1/2)}$ & 0.10 (0.092) &   &  -0.120 (-0.120)   \\
\hline
$A_{2}^{(1/2)}$ &0.69 (0.64) & 0.695 (0.680)   &      \\
\hline
$V_{}^{(1/2)}$ & 0.67 (0.61) & 0.695 (0.679)  &     \\
\hline
$A_{0}^{(3/2)}$ & -0.61 (-0.59) & 0.846 (0.834) &     \\
\hline
$A_{1}^{(3/2)}$ & -0.13 (-0.13)  &   &  -0.102 (-0.0956)   \\
\hline
$A_{2}^{(3/2)}$ & 0.72 (0.67) & 1.53 (1.54)  &      \\
\hline
$V_{}^{(3/2)}$ & -0.81 (-0.77) & 0.872 (0.873) &      \\
\hline
\end{tabular}
\end{center}
\end{table}

Regarding the $q^2$  dependance of the form factors, we find a very good agreement with
numerical results assuming  the following polar expression
\begin{equation}
F(q^2) =\frac{F(0)}{1-a\, \frac{q^2}{m_B^2} }\, ,
\end{equation}
the fitted values of $a$ can be found in Table \ref{t:fdfris2}. It is interesting to observe that
the effective pole mass is not far from the mass of the $B_c$ meson. A different $q^2$ dependence
exhibit the form factors $F_0$ and $A_1$; for them we use the form
\begin{equation}
F(q^2) =F(0)\left(1+ b\ q^2 \right)\, ,
\end{equation}
and the values of $b$ are collected in Table \ref{t:fdfris2}.

In conclusion we have obtained in a very simple constituent quark model all the
semileptonic form factors relevant to the transition of B into the low-lying
odd and even parity charmed mesons. The free parameters of the model have been
fixed by comparing model predictions with the $B\to D^\ast \ell \nu$ spectrum and
$B\to D \ell \nu$ branching ratio. Our numerical results are generally larger
than the results of other models. However, form factors reproduce the scaling laws
dictated by the HQET in the limit of infinitely heavy quark masses.

\appendix
\section{Form Factors $f_\pm$, $g$, $f$, $a_\pm$}
\label{a:0m0m1m} In this appendix we collect the analytical
expressions for the $0^- \to 0^-$ and $0^- \to 1^-$ form factors
defined in Eqs.~(\ref{e:F1F0}) and (\ref{e:gfapam}), respectively.

\begin{eqnarray}
f_+(q^2) & = & \int k^2 dk d
cos\theta\frac{\psi_I(k)\psi_F^\ast(k)}{ 8\,{\pi
}^2\,{m_{\CapitalIota }}\,
  {\sqrt{\left( {{d_{23}}}^2 - {{m_F}}^2 \right) \,
      \left( {{d_{12}}}^2 - {{m_{\CapitalIota }}}^2 \right) \,
      {{\CapitalEpsilon }_1}\,{{\CapitalEpsilon }_3}}}\,
  \left( {{m_F}}^2 - {{{\CapitalEpsilon }_F}}^2 \right)
}\times \nonumber\\
&&\left\{\frac{}{}( {d_{12}}\,{d_{23}}\,{{\CapitalEpsilon }_2} -
     {m_{\CapitalIota }}\,( {m_2}\,( -2\,{m_2} + {s_{13}} )  +
        {m_{\CapitalIota }}\,{{\CapitalEpsilon }_2} )  ) \,
   ( -{{m_F}}^2 + {{{\CapitalEpsilon }_F}}^2 )  + \right.\nonumber\\
&& \left.  k\,\qm\,\cos (\theta )\,(
{{\Mfunction{d}}_{12}}\,{d_{23}}\,
      ( {m_{\CapitalIota }} - {{\CapitalEpsilon }_F} )  +
     {{\Mfunction{m}}_{\CapitalIota }}\,
      ( -{{m_F}}^2 + {m_{\CapitalIota }}\,{{\CapitalEpsilon }_F} )
     )\frac{}{}\right\}\,,\\
f_-(q^2) & = &  \int k^2 dk dcos\theta
\frac{\psi_I(k)\psi_F^\ast(k)}{ 8\,{\pi }^2\,{m_{\CapitalIota }}\,
  {\sqrt{\left( {{d_{23}}}^2 - {{m_F}}^2 \right) \,
      \left( {{d_{12}}}^2 - {{m_{\CapitalIota }}}^2 \right) \,
      {{\CapitalEpsilon }_1}\,{{\CapitalEpsilon }_3}}}\,
  \left( {{m_F}}^2 - {{{\CapitalEpsilon }_F}}^2 \right)
} %
\times \nonumber\\
&& \left\{\frac{}{}( {d_{12}}\,{d_{23}}\,{{\CapitalEpsilon }_2} +
     {m_{\CapitalIota }}\,( {d_{13}}\,{m_2} +
        {m_{\CapitalIota }}\,{{\CapitalEpsilon }_2} )  ) \,
   ( -{{m_F}}^2 + {{{\CapitalEpsilon }_F}}^2 )  - \right.\nonumber\\
&& \left.   k\,\qm\,\cos (\theta )\,(
{{\Mfunction{d}}_{12}}\,{d_{23}}\,
      ( {m_{\CapitalIota }} + {{\CapitalEpsilon }_F} )  +
     {{\Mfunction{m}}_{\CapitalIota }}\,
      ( {{m_F}}^2 + {m_{\CapitalIota }}\,{{\CapitalEpsilon }_F} )
     )\frac{}{}\right\}\,,
\end{eqnarray}
\begin{eqnarray}
g(q^2) & = & \int k^2 dk d cos\theta \, \psi_I(k)\psi_F^\ast(k)\,
\frac{{m_2} + \frac{k^2\,{\sin (\theta )}^2}{{m_F} + {s_{23}}} +
    \frac{{d_{12}}\,{{\CapitalEpsilon }_2}}{{m_{\CapitalIota }}} -
    \frac{k\,\cos (\theta )\,\left( {{\Mfunction{d}}_{23}}\,
          {m_{\CapitalIota }} + {{\Mfunction{d}}_{12}}\,{{\CapitalEpsilon }_F}
         \right) }{\qm\,{m_{\CapitalIota }}}}{8\,{\pi }^2\,
    {\sqrt{\left( {{d_{23}}}^2 - {{m_F}}^2 \right) \,
        \left( {{d_{12}}}^2 - {{m_{\CapitalIota }}}^2 \right) \,
        {{\CapitalEpsilon }_1}\,{{\CapitalEpsilon }_3}}}}\,,\\
f(q^2) & = & \int k^2 dk d
cos\theta\frac{\psi_I(k)\psi_F^\ast(k)}{ 8\,{\pi }^2\,\left( {m_F}
+ {s_{23}} \right) \,
  {\sqrt{\left( {{d_{23}}}^2 - {{m_F}}^2 \right) \,
      \left( {{d_{12}}}^2 - {{m_{\CapitalIota }}}^2 \right) \,
      {{\CapitalEpsilon }_1}\,{{\CapitalEpsilon }_3}}}}\times \nonumber\\
&& \left\{ -( {{m_2}}^3\,{m_3} )  - k^2\,{m_1}\,{m_F} +
{m_1}\,{{m_F}}^3 +
  {{m_3}}^2\,{{m_{\CapitalIota }}}^2 + {m_3}\,{m_F}\,{{m_{\CapitalIota }}}^2 +
  {m_1}\,{{m_F}}^2\,{s_{23}} - \right.\nonumber\\
&& \left.  {{d_{23}}}^2\,{m_1}\,
   ( {m_F} + {s_{23}} )  +
  {d_{23}}\,{{m_1}}^2\,( {m_F} + {s_{23}} )  +
   k^2\,{m_{\CapitalIota }}\,{{\CapitalEpsilon }_F} -
  {{m_2}}^2\,( {m_F}\,( {m_3} + {m_F} )  +
     {{m_{\CapitalIota }}}^2 - 2\,{m_{\CapitalIota }}\,{{\CapitalEpsilon }_F}
     )  + \right.\nonumber\\
&& \left. k^2\,\cos (2\,\theta )\,
   ( {{\Mfunction{m}}_1}\,{m_F} - {{\Mfunction{m}}_2}\,{m_F} -
     {{\Mfunction{m}}_{\CapitalIota }}\,{{\CapitalEpsilon }_F} )  +
  {m_2}\,( k^2\,{m_F} + ( {m_3} - {m_F} ) \,
      {( {m_3} + {m_F} ) }^2 - \right.\nonumber\\
&& \left.  {m_F}\,{{m_{\CapitalIota }}}^2 +
     2\,( {m_3} + {m_F} ) \,{m_{\CapitalIota }}\,
      {{\CapitalEpsilon }_F} )
\right\}\,,\\
a_+(q^2) + a_-(q^2) & = & \int k^2 dk d
cos\theta\frac{\psi_I(k)\psi_F^\ast(k)}{ 4\,{\pi
}^2\,{\Mvariable{\qm}}^2\,{{m_{\CapitalIota }}}^2\,
  \left( {m_F} + {s_{23}} \right) \,
  {\sqrt{\left( {{d_{23}}}^2 - {{m_F}}^2 \right) \,
      \left( {{d_{12}}}^2 - {{m_{\CapitalIota }}}^2 \right) \,
      {{\CapitalEpsilon }_1}\,{{\CapitalEpsilon }_3}}}
      }\times\nonumber\\
&& \left\{ -( k^2\,{{m_F}}^2\,( {d_{12}}\,{m_F} -
       {m_{\CapitalIota }}\,{{\CapitalEpsilon }_F} )  )  +
  {\Mvariable{\qm}}^2\,{{\CapitalEpsilon }_2}\,
   ( 2\,{d_{12}}\,{m_F}\,{{\CapitalEpsilon }_2} +
     {m_{\CapitalIota }}\,( -{{m_3}}^2 +
        {( {m_2} + {m_F} ) }^2 - \right. \nonumber\\
&& \left.
        2\,{{\CapitalEpsilon }_2}\,{{\CapitalEpsilon }_F} )  )  +
  k\,\cos (\theta )\,( \Mfunction{k}\,\cos (\theta )\,
      ( {{\Mfunction{d}}_{12}}\,{m_F} -
        {{\Mfunction{m}}_{\CapitalIota }}\,{{\CapitalEpsilon }_F} ) \,
      ( {{{\Mfunction{m}}_F}}^2 + 2\,{{{\CapitalEpsilon }_F}}^2 )  + \right. \nonumber\\
&& \left.      \Mfunction{\qm}\,{{\CapitalEpsilon }_F}\,
      ( -4\,{d_{12}}\,{m_F}\,{{\CapitalEpsilon }_2} +
        {m_{\CapitalIota }}\,( {{m_3}}^2 -
           {( {m_2} + {m_F} ) }^2 +
           4\,{{\CapitalEpsilon }_2}\,{{\CapitalEpsilon }_F} )  )
     )
\right\}\,,\\
a_+(q^2) - a_-(q^2) & = & \int k^2 dk d
cos\theta\frac{\psi_I(k)\psi_F^\ast(k)}{ 4\,{\pi
}^2\,{\Mvariable{\qm}}^2\,{m_{\CapitalIota }}\,
  \left( {m_F} + {s_{23}} \right) \,
  {\sqrt{\left( {{d_{23}}}^2 - {{m_F}}^2 \right) \,
      \left( {{d_{12}}}^2 - {{m_{\CapitalIota }}}^2 \right) \,
      {{\CapitalEpsilon }_1}\,{{\CapitalEpsilon }_3}}}\,{{\CapitalEpsilon }_F}
}\times \nonumber\\
&& \left\{ {m_{\CapitalIota }}\,( -( k^2\,
        ( {\Mvariable{\qm}}^2 + {{m_F}}^2 )  )  -
     {\Mvariable{\qm}}^2\,( {m_2}\,( {m_F} + {s_{23}} )  -
        {m_{\CapitalIota }}\,{{\CapitalEpsilon }_2} )  ) \,
   {{\CapitalEpsilon }_F} + {d_{12}}\,
   ( k^2\,{m_F}\,( {\Mvariable{\qm}}^2 + {{m_F}}^2 )  - \right. \nonumber\\
&& \left. {\Mvariable{\qm}}^2\,( {m_F} + {s_{13}} ) \,
      {{\CapitalEpsilon }_2}\,{{\CapitalEpsilon }_F} )  +
  k\,\cos (\theta )\,( \Mfunction{k}\,\cos (\theta )\,
      ( -( {{\Mfunction{d}}_{12}}\,{m_F} )  +
        {{\Mfunction{m}}_{\CapitalIota }}\,{{\CapitalEpsilon }_F} ) \,
      ( {\Mfunction{\qm}}^2 + {{m_F}}^2 + 2\,{{{\CapitalEpsilon }_F}}^2
        )  +\right. \nonumber\\
&& \left.  \Mfunction{\qm}\,{{\CapitalEpsilon }_F}\,
      ( {d_{12}}\,( 2\,{m_F}\,{{\CapitalEpsilon }_2} +
           ( {m_F} + {s_{13}} ) \,{{\CapitalEpsilon }_F} )  +
        {m_{\CapitalIota }}\,( {d_{23}}\,
            ( {m_F} + {s_{23}} )  -
           ( {m_{\CapitalIota }} + 2\,{{\CapitalEpsilon }_2} ) \,
            {{\CapitalEpsilon }_F} )  )  )
\right\}\,,
\end{eqnarray}
where $d_{ij} = m_i-m_j$ and $s_{ij} = m_i+m_j$ (with $m_i$ the mass of $i-$quark);
$m_I$ and $m_F$ are the masses of the initial and final mesons, respectively. $E_{F} (=
\sqrt{\qv^{\, 2} + m_{F}^2})$ represents the energy of the final meson.
The angle $\theta$ is defined in Section \ref{s:DCFDF} after Eq. (\ref{e:f}).

\section{Form Factors $G^{(\prime)}$, $F^{(\prime)}$, $A^{(\prime)}_\pm$}
\label{a:0m1p}

In this appendix we give the expressions of the form factors
appearing in Eq. (\ref{e:gfapam1P1}) ($0^-\to ^1\!\!\!P_1$
transitions). We use the same notations of the previous appendix.
\begin{eqnarray}
G(q^2) & = & -\int k^2 dk d cos\theta
\frac{ \psi_I(k)
\psi^\ast_F(k){\sqrt{3}}\,k^2\,{\sin (\theta )}^2{m_F}}
  {8{\pi }^2\,\left( -{{d_{23}}}^2 + {{m_F}}^2 \right)
    {\sqrt{\left( -{{d_{12}}}^2 + {{m_{\CapitalIota }}}^2 \right)
    \left( {{m_F}}^2 - {{s_{23}}}^2 \right) {{\CapitalEpsilon }_1}
        {{\CapitalEpsilon }_3}}}}
        \\
F(q^2) & = &  -\int k^2 dk d cos\theta \psi_I(k)\,
\psi^\ast_F(k)
\frac{{\sqrt{3}}\,k^2\,{\sin (\theta )}^2\,{m_F}\,
    \left( -q^2 + 2\,{d_{12}}\,{d_{23}} + {{m_F}}^2 +
      {{m_{\CapitalIota }}}^2 \right) }{8\,{\pi }^2\,
    \left( {{d_{23}}}^2 - {{m_F}}^2 \right) \,
    {\sqrt{\left( -{{d_{12}}}^2 + {{m_{\CapitalIota }}}^2 \right) \,
        \left( {{m_F}}^2 - {{s_{23}}}^2 \right) \,{{\CapitalEpsilon }_1}\,
        {{\CapitalEpsilon }_3}}}}
\end{eqnarray}
\begin{eqnarray}
A_+(q^2)+A_-(q^2) &=& -\int k^2 dk d cos\theta
                      \frac{\psi_I(k) \psi^\ast_F(k)\,\sqrt{3}\, m_F}
                      {8\,{\pi }^2\,{\Mvariable{\qm}}^2\left( -{{d_{23}}}^2 + {{m_F}}^2 \right)
  {{m_{\CapitalIota }}}^2{\sqrt{\left( {{d_{12}}}^2 -
        {{m_{\CapitalIota }}}^2 \right)
      \left( -{{m_F}}^2 + {{s_{23}}}^2 \right) {{\CapitalEpsilon }_1}
      {{\CapitalEpsilon }_3}}}}\times \nonumber\\
&&\left\{
k^2\,{\sin (\theta )}^2\,{{m_F}}^2\,
   \left( -q^2 + 2\,{d_{12}}\,{d_{23}} + {{m_F}}^2 +
     {{m_{\CapitalIota }}}^2 \right)  -
  2\,\left( \Mvariable{\qm}\,{{\CapitalEpsilon }_2} -
     k\,\cos (\theta )\,{{\Mfunction{\CapitalEpsilon }}_F} \right) \times \right.\nonumber\\
&&\left.
   \left( \Mvariable{\qm}\,\left( \left( {{d_{23}}}^2 - {{m_F}}^2 \right) \,
         {m_{\CapitalIota }} + 2\,{d_{12}}\,{d_{23}}\,{{\CapitalEpsilon }_2} +
        \left( -q^2 + {{m_F}}^2 + {{m_{\CapitalIota }}}^2 \right) \,
         {{\CapitalEpsilon }_2} \right)  -\right.\right.\nonumber\\
&&\left.\left.
     k\,\cos (\theta )\,\left( -{\Mfunction{q}}^2 + 2\,{d_{12}}\,{d_{23}} +
        {{m_F}}^2 + {{{\Mfunction{m}}_{\CapitalIota }}}^2 \right) \,
      {{\Mfunction{\CapitalEpsilon }}_F} \right)
\frac{}{}\right\}
\end{eqnarray}
\begin{eqnarray}
A_+(q^2)-A_-(q^2) &=& -\int
                      \frac{k^2\, dk\, d cos(\theta)\,\psi_I(k) \psi^\ast_F(k) \,\sqrt{3}\, m_F}{4\,{\pi }^2
                      {\Mvariable{\qm}}^2\,\left( {{d_{23}}}^2 - {{m_F}}^2 \right)
  {m_{\CapitalIota }}\left( -q^2 + {{m_F}}^2 + {{m_{\CapitalIota }}}^2 \right)
    {\sqrt{\left( {{d_{12}}}^2 - {{m_{\CapitalIota }}}^2 \right)
        \left( -{{m_F}}^2 + {{s_{23}}}^2 \right) {{\CapitalEpsilon }_1}
      {{\CapitalEpsilon }_3}}}}\times\nonumber\\
&& \left\{\frac{}{}
2\,{m_{\CapitalIota }}\,{{\CapitalEpsilon }_F}\,
  \left( {\Mvariable{\qm}}^2\,\left( {{d_{12}}}^2 - {{m_{\CapitalIota }}}^2
       \right) \,{{\CapitalEpsilon }_2} +
    k^2\,{{\CapitalEpsilon }_F}\,
     \left( {d_{12}}\,{d_{23}} +
       {m_{\CapitalIota }}\,{{\CapitalEpsilon }_F} \right)  +\right.\right.\nonumber\\
&&\left.\left.
    k\,\cos (\theta )\,\left( -3\,\Mfunction{k}\,\cos (\theta )\,
        {{\Mfunction{\CapitalEpsilon }}_F}\,
        \left( {d_{12}}\,{d_{23}} +
          {m_{\CapitalIota }}\,{{\CapitalEpsilon }_F} \right)  + \right.\right.\right.\nonumber\\
&&\left.\left.\left.
       \Mfunction{\qm}\,\left( {{m_{\CapitalIota }}}^2\,{{\CapitalEpsilon }_F} +
          2\,{m_{\CapitalIota }}\,{{\CapitalEpsilon }_2}\,
           {{\CapitalEpsilon }_F} +
          {d_{12}}\,\left( 2\,{d_{23}}\,{{\CapitalEpsilon }_2} -
             {d_{12}}\,{{\CapitalEpsilon }_F} \right)  \right)  \right)  \right)
\frac{}{}\right\}
\end{eqnarray}

Regarding the form factors in Eq. (\ref{e:gfapam3P1}) we have
\begin{eqnarray}
G^\prime(q^2) &=& -\sqrt{\frac{3}{2}} \int  \frac{k^2 dk d cos\theta\, \psi_I(k)
\psi^\ast_F(k)}{8{\pi }^2\Mvariable{\qm}\left( -{{d_{23}}}^2 + {{m_F}}^2 \right)
  {m_{\CapitalIota }}{\sqrt{\left( {{d_{12}}}^2 - {{m_{\CapitalIota }}}^2
        \right) \left( -{{m_F}}^2 + {{s_{23}}}^2 \right)
      {{\CapitalEpsilon }_1}{{\CapitalEpsilon }_3}}}} \times\nonumber\\
&& \left\{\frac{}{}
-\left( \Mvariable{\qm}\,\left( \left( {d_{23}}\,
           \left( k^2 + {d_{23}}\,{m_2} \right)  - {m_2}\,{{m_F}}^2 \right) \,
        {m_{\CapitalIota }} + {d_{12}}\,
        \left( {{d_{23}}}^2 - {{m_F}}^2 \right) \,{{\CapitalEpsilon }_2} \right)
        \right)  +  \right.\nonumber\\
&& \left. k\,\cos (\theta )\,
   \left( \Mfunction{k}\,\Mvariable{\qm}\,\cos (\theta )\,
      {{\Mfunction{d}}_{23}}\,{m_{\CapitalIota }} +
     \left( {{{\Mfunction{d}}_{23}}}^2 - {{m_F}}^2 \right) \,
      \left( {{\Mfunction{m}}_{\CapitalIota }}\,{s_{23}} +
        {{\Mfunction{d}}_{12}}\,{{\CapitalEpsilon }_F} \right)  \right)\frac{}{}
\right\}
\end{eqnarray}
\begin{eqnarray}
F^\prime(q^2) &=& -\sqrt{\frac{3}{2}}\int  \frac{k^2 dk d cos\theta\, \psi_I(k)
\psi^\ast_F(k)}{ 8\, {\pi }^2\,\left( -{{d_{23}}}^2 + {{m_F}}^2 \right)
  {\sqrt{\left( {{d_{12}}}^2 - {{m_{\CapitalIota }}}^2 \right)
      \left( -{{m_F}}^2 + {{s_{23}}}^2 \right) {{\CapitalEpsilon }_1}
      {{\CapitalEpsilon }_3}}}}\times\nonumber\\
&& \left\{
-2\,k^2\,{\sin (\theta )}^2\,\left( {d_{12}}\,{{m_F}}^2 +
     {d_{23}}\,{m_{\CapitalIota }}\,{{\CapitalEpsilon }_F} \right)  +
  \left( {{d_{23}}}^2 - {{m_F}}^2 \right) \,
   \left( -\left( {{m_1}}^2\,{s_{23}} \right)  +
     {{m_{\CapitalIota }}}^2\,{s_{23}} +
     {m_1}\,\left( -{{m_F}}^2 + {{s_{23}}}^2 \right)  + \right.\right. \nonumber\\
&&\left.\left.
     {m_2}\,\left( {{m_F}}^2 - {m_3}\,{s_{23}} -
        2\,{m_{\CapitalIota }}\,{{\CapitalEpsilon }_F} \right)  \right)
\right\}
\end{eqnarray}

\begin{eqnarray}
A_+^\prime(q^2) + A_-^\prime(q^2) &=& -\sqrt{\frac{3}{2}} \int
\frac{k^2 dk d cos\theta\,\psi_I(k) \psi^\ast_F(k)}{16{\pi }^2{\Mvariable{\qm}}^2\,
\left( -{{d_{23}}}^2 + {{m_F}}^2 \right)
  {{m_{\CapitalIota }}}^2{\sqrt{\left( {{d_{12}}}^2 -
        {{m_{\CapitalIota }}}^2 \right) \,
      \left( -{{m_F}}^2 + {{s_{23}}}^2 \right) {{\CapitalEpsilon }_1}
      {{\CapitalEpsilon }_3}}}
}\times\nonumber\\
&& \left\{
4\,\left( {\Mvariable{\qm}}^2\,{{\CapitalEpsilon }_2}\,
     \left( {{m_F}}^2\,{m_{\CapitalIota }}\,{s_{23}} +
       2\,{d_{12}}\,{{m_F}}^2\,{{\CapitalEpsilon }_2} +
       {d_{23}}\,{m_{\CapitalIota }}\,
        \left( -\left( {d_{23}}\,{s_{23}} \right)  +
          2\,{{\CapitalEpsilon }_2}\,{{\CapitalEpsilon }_F} \right)  \right)  - \right.\right.\nonumber\\
&&\left.\left.
    k\,\Mvariable{\qm}\,\cos (\theta )\,{{\Mfunction{\CapitalEpsilon }}_F}\,
     \left( {{m_F}}^2\,{m_{\CapitalIota }}\,{s_{23}} +
       4\,{d_{12}}\,{{m_F}}^2\,{{\CapitalEpsilon }_2} +
       {d_{23}}\,{m_{\CapitalIota }}\,
        \left( -\left( {d_{23}}\,{s_{23}} \right)  +
          4\,{{\CapitalEpsilon }_2}\,{{\CapitalEpsilon }_F} \right)  \right)  + \right.\right.\nonumber\\
&&\left.\left.
    k^2\,\left( {d_{12}}\,{{m_F}}^2 +
       {d_{23}}\,{m_{\CapitalIota }}\,{{\CapitalEpsilon }_F} \right) \,
     \left( -{{m_F}}^2 + {\cos (\theta )}^2\,
        \left( {{m_F}}^2 + 2\,{{{\CapitalEpsilon }_F}}^2 \right)  \right)
    \right)
\frac{}{}
\right\}\\
A_+^\prime(q^2) - A_-^\prime(q^2) &=& + \sqrt{\frac{3}{2}} \int
\frac{k^2\, dk\, d cos(\theta)\, \psi_I(k) \psi^\ast_F(k)}
{8\,{\pi }^2\,{\Mvariable{\qm}}^2\,\left( -{{d_{23}}}^2 + {{m_F}}^2 \right)
  {m_{\CapitalIota }}\, {{\CapitalEpsilon }_F}\,
   {\sqrt{\left( {{d_{12}}}^2 - {{m_{\CapitalIota }}}^2 \right)
      \left( -{{m_F}}^2 + {{s_{23}}}^2 \right) {{\CapitalEpsilon }_1}
      {{\CapitalEpsilon }_3}}}
}\times\nonumber\\
&& \left\{
-\left( k^2\,\left( {d_{12}}\,{{m_F}}^2 +
       {d_{23}}\,{m_{\CapitalIota }}\,{{\CapitalEpsilon }_F} \right) \,
     \left( -{{m_F}}^2 + 2\,{{{\CapitalEpsilon }_F}}^2 \right)  \right)  +
  {\Mvariable{\qm}}^2\,\left( 2\,{{d_{23}}}^2\,{m_2}\,{m_{\CapitalIota }}\,
      {{\CapitalEpsilon }_F} + {d_{23}}\,
      \left( k^2\,{m_{\CapitalIota }} + \right.\right.\right.\nonumber\\
&&\left.\left.\left.
        2\,\left( {{m_1}}^2 + {m_2}\,{m_3} - {{m_{\CapitalIota }}}^2 -
           {m_1}\,{s_{23}} \right) \,{{\CapitalEpsilon }_2} \right) \,
      {{\CapitalEpsilon }_F} + {{m_F}}^2\,
      \left( k^2\,{d_{12}} - 2\,\left( {m_2}\,{m_{\CapitalIota }} +
           {d_{12}}\,{{\CapitalEpsilon }_2} \right) \,{{\CapitalEpsilon }_F}
        \right)  \right)  + \right.\nonumber\\
&& \left.
k\,{{\CapitalEpsilon }_F}\,
   \left( -3\,k\,\cos (2\,\theta )\,{{\Mfunction{\CapitalEpsilon }}_F}\,
      \left( {d_{12}}\,{{m_F}}^2 +
        {d_{23}}\,{m_{\CapitalIota }}\,{{\CapitalEpsilon }_F} \right)
        - \right.\right.\nonumber\\
&& \left.\left.
     2\,\Mvariable{\qm}\,\cos (\theta )\,
      \left( {{{\Mfunction{d}}_{23}}}^2\,{m_{\CapitalIota }}\,{s_{23}} -
        {{{\Mfunction{m}}_F}}^2\,
         \left( {m_{\CapitalIota }}\,{s_{23}} +
           2\,{d_{12}}\,{{\CapitalEpsilon }_2} \right)  + \right.\right.\right.\nonumber\\
&&\left.\left.\left.
        {{\Mfunction{d}}_{23}}\,{{m_1}}^2\,{{\CapitalEpsilon }_F} +
        \left( {{{\Mfunction{m}}_2}}^2\,{m_3} -
           {{\Mfunction{m}}_1}\,
            \left( {{m_F}}^2 + {d_{23}}\,{s_{23}} \right)  +
           {{\Mfunction{m}}_3}\,{m_{\CapitalIota }}\,
            \left( {m_{\CapitalIota }} + 2\,{{\CapitalEpsilon }_2} \right)  -
            \right.\right.\right.\right.\nonumber\\
&&\left.\left.\left.\left.
           {{\Mfunction{m}}_2}\,\left( {{m_3}}^2 - {{m_F}}^2 +
              {{m_{\CapitalIota }}}^2 +
              2\,{m_{\CapitalIota }}\,{{\CapitalEpsilon }_2} \right)  \right) \,
         {{\Mfunction{\CapitalEpsilon }}_F} \right)  \right)\frac{}{}
\right\}
\end{eqnarray}

\end{document}